\documentclass[aps,prd,twocolumn,groupedaddress,superscriptaddress,preprintnumbers,nofootinbib]{revtex4}
\usepackage{amsmath,amssymb,epsfig,ifthen,capt-of,calc}
\usepackage{color}

\newcommand{\eq}[1]{Eq.~(\ref{#1})}

\newcommand{\FTD}{F_{\rm TD}}

\newcommand{\ev}{\,{\rm eV}}
\newcommand{\kev}{\,{\rm keV}}
\newcommand{\mev}{\,{\rm MeV}}
\newcommand{\gev}{\,{\rm GeV}}

\begin{document}
\preprint{PNUTP-11-A01}
\title{
\bf Techni-dilaton as Dark Matter}

\author{Ki-Young Choi~\footnote{
K.Y.C. is currently at Asia Pacific Center for Theoretical Physics,  POSTECH, Korea. 
}}
\affiliation{ Department of Physics,  
                   Pusan National University, Busan 609-735, Korea.}
\author{Deog Ki Hong}
\affiliation{ Department of Physics,  
                   Pusan National University, Busan 609-735, Korea.}
\affiliation{Asia Pacific Center for Theoretical Physics,  POSTECH, Pohang 709-784, Korea.}  
 \author{Shinya Matsuzaki~\footnote{
S.M. is currently at Maskawa Institute for Science and Culture, Kyoto Sangyo University, Japan. 
}} 
\affiliation{ Department of Physics,  
                   Pusan National University, Busan 609-735, Korea.}    
                                 
\date{\today}

\begin{abstract}
We propose a new dark matter candidate, {\it light decoupled techni-dilaton}, 
which arises from the almost scale-invariant/conformal (extreme walking) technicolor. 
We investigate its characteristic nature and discuss several cosmological and astrophysical constraints. 
It turns out that techni-dilatons are extremely weakly interacting and 
produced dominantly by the non-thermal mechanism to become the main component of dark matter 
with mass range between around 0.01 eV and 500 eV for extreme walking technicolor scenarios. 
\end{abstract}
\maketitle

\section{Introduction}

The recent observation in astrophysics and cosmology clearly indicates  dark matter constitutes 
23\% of total energy  in the present Universe.
The combined analysis, including galaxy rotation curves, gravitational lensing, cosmic microwave background and structure formation, 
shows that dark matter is non-baryonic, weakly interacting and cold. 
It therefore demands to go beyond the standard model (SM) of particle physics  to accommodate  dark matter.

Certain viable models beyond the SM involve scenarios of dynamical electroweak symmetry breaking, so-called 
technicolor (TC)~\cite{Weinberg:1975gm}, in which the electroweak symmetry is broken dynamically without a fundamental  Higgs, 
by the nonperturbative condensate of techni-fermion and anti-techni-fermion, 
$\langle \bar{F}F \rangle $, triggered by the new gauge interaction. 
The electroweak precision data and the absence of flavor changing neutral currents constrain properties of  
the TC gauge dynamics~\cite{Farhi:1980xs} to be 
almost conformal or scale-invariant (walking)~\cite{Holdom:1981rm,Yamawaki:1985zg,Akiba:1985rr,Hong:2004td}, 
characterized by a large anomalous dimension ($\gamma_m \simeq 1$) of techni-fermion bilinear $\bar{F}F$. 
The TC gauge coupling ($\alpha$) then has an almost nonrunnning/conformal behavior  between 
the techni-fermion mass scale $m_F ={\cal O}(10^3\, {\rm GeV})$ at infrared (IR) and 
the  intrinsic scale of walking TC at ultraviolet (UV), $\Lambda_{\rm TC}$, which is usually identified with 
extended TC (ETC) scale~\cite{Dimopoulos:1979es}, 
$\Lambda_{\rm ETC}= {\cal O}(10^6\, {\rm GeV})$, but not in our proposal.

The almost conformal property implies existence of the approximate scale invariance, 
whose spontaneous breaking then leads to generation of (almost) massless dilaton.  
In walking TC the (approximate) scale symmetry is broken by the nonperturbative generation of 
techni-fermion mass $m_F$.  
The dynamical generation of $m_F$ is characterized by essential singularity scaling, 
what is called Miransky scaling~\cite{Miransky:1984ef,Hong:1989zza,Kaplan:2009kr} 
tied with the conformal phase transition~\cite{Miransky:1996pd}: 
\begin{equation}
m_F\simeq \Lambda_{\rm TC}\,e^{-\pi/\sqrt{\alpha_m/\alpha_{\rm cr}-1}}\,,
\label{miransky}
\end{equation} 
where $\alpha_{\rm cr}$ is the critical coupling for the chiral symmetry breaking and $\alpha_m\,(>\alpha_{\rm cr})$ is the gauge coupling measured at $m_F$.  
By the Goldstone theorem, therefore, dilaton, the Nambu-Goldstone boson for the spontaneously broken  scale symmetry,  
emerges in walking TC~\cite{Bardeen:1985sm,Yamawaki:1985zg}.

Actually, the ``fixed" coupling $\alpha_m$ becomes dependent of $\Lambda_{\rm TC}/m_F$ and starts ``running" 
according to Eq.(\ref{miransky}), to break the scale symmetry {\it explicitly}: 
\begin{eqnarray}  
\partial_\mu D^\mu &=& \frac{\beta(\alpha)}{4 \alpha^2} 
\langle \alpha G_{\mu\nu}^2 \rangle 
\,, 
\nonumber \\ 
\beta(\alpha) &=& \Lambda_{\rm TC} \frac{\partial \alpha}{\partial \Lambda_{\rm TC}} 
= - \frac{2 \alpha_{\rm cr}}{\pi} \left( \frac{\alpha}{\alpha_{\rm cr}} - 1 \right)^{3/2} 
\,,  \label{NPbeta}
\end{eqnarray} 
where $D_\mu$ is the dilatation current composed of TC sector fields 
and $G_{\mu\nu}$ is the field strength tensor of techni-gluons. 
Thus the nonperturbative generation of techni-fermion mass {\it explicitly} (as well as spontaneously) 
breaks the scale symmetry due to the nonperturbative running (walking) given by Miransky scaling, Eq.(\ref{miransky}). 
The dilaton therefore becomes massive (pseudo Nambu-Goldstone boson) 
by this ``nonperturbative-explicit breaking effect", 
which was long ago predicted in walking TC as a techni-dilaton (TD)~\cite{Yamawaki:1985zg}. 
Thereby, {\it the TD mass $M_{\rm TD}$ and the coupling, set by the decay constant $F_{\rm TD}$,  
are associated to the nonperturbative beta function $\beta(\alpha)$}. 
This can be easily seen if we assume the partially conserved dilatation currents (PCDCs): 
\begin{equation} 
F_{\rm TD}^2 M_{\rm TD}^2 
= - 16 \,{\cal E}_{\rm vac} =  - \frac{\beta(\alpha)}{\alpha^2} \langle \alpha G_{\mu\nu}^2 \rangle 
\,,  \label{PCDC:original}
\end{equation} 
where ${\cal E}_{\rm vac}$ denotes the vacuum energy density governed by the techni-gluon condensation triggered by 
the nonperturbative generation of $m_F$.

The nonperturbative beta function $\beta(\alpha)$ in Eq.(\ref{NPbeta}) describes 
the running (walking) for a wide range from the (UV) intrinsic scale $\Lambda_{\rm TC}$ to 
the (IR) techni-fermion mass scale $m_F$, 
which can be {\it extremely hierarchical}, say, $\Lambda/m_F \approx 10^9$ 
due to the essential singularity scaling Eq.(\ref{miransky}). 
For instance,  if  $\alpha_m=1.03\,\alpha_{\rm cr}$, 
then $\Lambda_{\rm TC}\approx 10^9\, m_F$ and 
the theory is {\it extremely walking}. 
On the other hand, if $\alpha_m=1.25\,\alpha_{\rm cr}$, 
one would have $\Lambda_{\rm TC}\approx 10^3\,m_F$, 
close to $\Lambda_{\rm ETC}$.  
Though the precise value of $\alpha_m$ is determined dynamically,  
it lies parametrically between $\alpha_{\rm cr}$ and $\alpha_*$, 
the quasi IR fixed point of walking TC~\cite{Caswell:1974gg}.  
Thus the {\it extremely walking} theory which yields the {\it extreme} scale hierarchy $\Lambda_{\rm TC}/m_F \approx 10^9$ 
can be realized if $\alpha_{\rm cr}$ and $\alpha_*$ are very close to each other.

Since the TD mass $M_{\rm TD}$ and decay constant $F_{\rm TD}$ couple directly with $\beta(\alpha)$ 
(via the nonperturbative scale anomaly Eq.(\ref{PCDC:original})) and hence originate from either 
the UV scale ($\Lambda_{\rm TC}$)~\footnote{
It could be the scale $\mu_{\rm cr}$ at which $\alpha(\mu_{\rm cr}) =\alpha_{\rm cr}$ 
instead of $\Lambda_{\rm TC}$. 
But, we do not differentiate these two UV scales here, 
because there is no large hierarchy between them~\cite{Hashimoto:2010nw}.} or the IR scale ($m_F$), 
they could be of the same order, or even {\it extremely hierarchical} as well as $\Lambda_{\rm TC}$ and $m_F$. 
However, recall that the TD should be lighter than other TC hadrons like techni-rho meson 
(with mass of order of a few TeV) because of the pseudo Nambu-Goldstone boson's nature. 
The scale of $M_{\rm TD}$ should therefore be set by the IR scale, $m_F$, which leaves us only 
two possibilities that i) $M_{\rm TD} \sim {\cal O}(m_F)$ or ii) $M_{\rm TD} \ll {\cal O}(m_F)$.

Several attempts to calculate $M_{\rm TD}$ have so far been 
performed in a different context based on straightforward nonperturbative calculations with 
some approximations assumed~\cite{Shuto:1989te,Bardeen:1991sv,Carena:1992cg,Harada:2003dc}. 
Those results suggest that $M_{\rm TD} \sim {\cal O}(m_F)$ near the criticality of walking TC ($\alpha \approx \alpha_{\rm cr}$) 
and hence favor the case i). 
On the other hand, a recent holographic analysis~\cite{Haba:2010hu} implies that 
$M_{\rm TD}/m_F \ll 1$ near the criticality, which supports the case ii).   
Thus the hierarchy between $M_{\rm TD}$ and $m_F$ is still controversial at present, 
so one cannot exclude either the case i) or the case ii) above.

As for the TD decay constant $F_{\rm TD}$, no explicit estimate based on nonperturbative analysis has been 
carried out.  
Recently, however, one implication from holography~\cite{Haba:2010hu} has been given 
to suggest that $F_{\rm TD}\gg m_F$ near the criticality, 
so one expects $F_{\rm TD} \sim \Lambda_{\rm TC}$ 
from the nonperturbative scale anomaly (Eqs.(\ref{NPbeta})-(\ref{PCDC:original})). 
The presence of such an extremely large $F_{\rm TD}$ might be reasonable  
from a point of view of the naive dimensional analysis~\cite{Manohar:1983md} 
since $F_{\rm TD}\sim \Lambda_{\rm TC}/4\pi$ when one considers {\`a} la chiral 
Lagrangian describing the dilaton dynamics, analogously to QCD pions.

Indeed, it turns out that the PCDC relation Eq.(\ref{PCDC:original}) allows the {\it extremely hierarchical} 
scales between $M_{\rm TD}$ and $F_{\rm TD}$:  
\begin{equation} 
  \frac{M_{\rm TD}}{m_F} \ll 1 
  \qquad 
  {\rm when}
  \qquad  
    \frac{F_{\rm TD}}{m_F} \gg 1   
\, , \label{decoupledTD}
\end{equation} 
which can be clarified as follows. 
The estimate of the vacuum energy density ${\cal E}_{\rm vac}$ in Eq.(\ref{PCDC:original}) 
has been performed with some assumptions~\cite{Appelquist:2010gy} 
or approximations~\cite{Miransky:1989qc,Hashimoto:2010nw} independently of calculation of $M_{\rm TD}$. 
In an {\it extremely walking} case such as when $\Lambda_{\rm TC}/m_F \approx 10^9$  
the nonrunning approximation for the TC gauge coupling (what is called standing limit)  
is sufficient for evaluating the vacuum energy. 
We then have ${\cal E}_{\rm vac}=- (N_{\rm TC} N_{\rm TF}/{\pi^4}) m_F^4$~\cite{Miransky:1989qc} and hence   
\begin{equation} 
  F_{\rm TD}^2 M_{\rm TD}^2 = \frac{16 N_{\rm TC} N_{\rm TF}}{\pi^4} m_F^4 \, , \label{PCDC}
\end{equation}
where $N_{\rm TC}$ and $N_{\rm TF}$ respectively stand for the number of TC and that of techni-fermions. 
Thus the PCDC Eq.(\ref{PCDC}) is consistent with the {\it extremely hierarchical} scenario Eq.(\ref{decoupledTD}).

Since the walking TC yields $\gamma_m \simeq 1$ for the techni-fermion bilinear operator $\bar{F}F$, 
the induced four-fermi operator $(\bar{F}F)^2$ having ${\rm dim}(\bar{F}F)^2 \simeq 4$ becomes marginal as well as the TC gauge coupling $\alpha$ 
in the sense of renormalization group analysis. 
The form of scale anomaly Eq.(\ref{NPbeta}), hence the PCDC relation in Eqs.(\ref{PCDC:original}) and (\ref{PCDC}), should then 
be modified by the presence of the four-fermi interaction. 
One might therefore think that our observation based on Eq.(\ref{PCDC}) would then make no sense. 
Actually, such four-fermi effects have been intensively studied through the analysis on the 
planar QED with nonrunning gauge coupling and four-fermion interactions added 
(what is called gauged Nambu-Jona-Lasinio (NJL) model)~\cite{Bardeen:1985sm,Nonoyama:1989dq,Shuto:1989te,Leung:1989hw,Bardeen:1991sv,Carena:1992cg}. 
Particularly in Refs.~\cite{Nonoyama:1989dq,Shuto:1989te}, 
the vacuum energy density ${\cal E}_{\rm vac}$ was explicitly computed in the gauged NJL model 
with the nonrunning gauge coupling, so that the result essentially remains the same as in Eq.(\ref{PCDC}), 
${\cal E}_{\rm vac} \sim N_{\rm TC}N_{\rm TF} m_F^4$. 
Thus our observation above will be unaffected even in the presence of the marginal four-fermi operator~\footnote{ 
Even for a perturbatively running case, it leads 
to essentially the same result on ${\cal E}_{\rm vac}$ as that in Eq.(\ref{PCDC})~\cite{Hashimoto:2010nw} 
within a 5\% uncertaity. 
This reflects the fact that the mass and coupling of TD are tied to the nonperturbative scale anomaly (\ref{NPbeta}) which 
has nothing to do with how the theory is perturbatively (fully or almost) scale invariant, as long as the 
dynamical fermion mass is generated in accord with {\`a} la Miransky scaling (\ref{miransky}). 
}.

It is worth exploring what would happen if the {\it extremely hierarchical} scenario Eq.(\ref{decoupledTD}) 
could be realized~\footnote{ An alternative non-hierarchical scenario ($M_{\rm TD} \sim F_{\rm TD} \sim m_F$) 
is of course not excluded by Eq.(\ref{PCDC}), which would be relevant to 
the TD LHC physics~\cite{Hashimoto:2011ma}. }.   
In that case, the TD could become almost massless ($\ll m_F$) 
with the extremely large decay constant $F_{\rm TD}$ ($\sim \Lambda_{\rm TC}) \gg m_F \gg M_{\rm TD}$. 
The TD interactions with other particles are then highly suppressed by the $\FTD$, 
which would make the TD decoupled from the thermal background in the early Universe. 
Such a TD can therefore be a dark matter candidate, that we shall call a {\it light decoupled TD}.

In this paper we propose the light decoupled TD 
as a new candidate for dark matter which arises from 
{\it extreme walking} TC with $\Lambda_{\rm TC} (\sim F_{\rm TD}) \approx 10^9~m_F$~\footnote{
To make the scenario of light decoupled TD 
phenomenologically viable and in particular 
accommodate the SM fermion masses, 
one may introduce 
an extra gauge interaction communicating the walking TC and SM sectors, 
which is like an ETC, 
around at the scale $\simeq 10^6$ GeV below the intrinsic scale $\Lambda_{\rm TC}$. }. 
It has actually been suggested in Ref.~\cite{Hashimoto:2010nw} without explicit estimate 
 that the decoupled TD might be a candidate for dark matter. 
In the following we show that the light decoupled TD can certainly 
be a new candidate for the dark matter by explicitly estimating the lifetime and relic abundance, 
consistently with cosmological and astrophysical constraints.

We describe here all the essential features of TD as dark matter in a self-contained manner, 
though some of the detailed calculations are deferred to a forthcoming longer version, 
which deals with other issues as well~\cite{longer-version}.

\section{Lifetime}

The TD decays into two photons through the scale anomaly 
involving techni-fermions ($F$) and SM fermions ($f$) in the loop. 
   For the light decoupled TD case with $M_{\rm TD} \ll m_F, m_f$, 
the decay rate $\Gamma(D \to \gamma\gamma)$ is given as 
\begin{equation} 
 \Gamma(D \to \gamma\gamma)
\simeq 
 \frac{\alpha_{\rm EM}^2}{36 \pi^3} \frac{M_{\rm TD}^3}{F_{\rm TD}^2}  |{\cal C}|^2
 \,, \label{decay-width2}
\end{equation} 
where $\alpha_{\rm EM}$ denotes the fine structure constant and 
 ${\cal C} \equiv \sum_{f'=f,F} N_{\rm TC}^{f'} N_c^{f'} Q_{f'}^2 $.
Here $Q_{f'}$ is the electromagnetic charge of $f'$ and
$N_{\rm TC}^{f'}=N_{\rm TC}$ (or 1) for $f'=F$ (or $f$)
while  $N_c^{f'}=3$ (or $1$) for fermions belonging to fundamental
representation (or singlet) of the QCD color. 
 For the SM alone, we have ${\cal C}_{\rm SM}=8$ and 
one-family TC with $N_{\rm TC}=2$ and $N_{\rm TF}=8$ gives ${\cal C}_{\rm one-family}=8$, so that 
${\cal C}={\cal C}_{\rm SM}+{\cal C}_{\rm one-family}=16$. 
Using \eq{PCDC}, we estimate the lifetime of TD, $\tau_{\rm TD}$, to get  
\begin{equation} 
 \tau_{\rm TD} \simeq 10^{17} \, {\rm sec} \, (N_{\rm TC}N_{\rm TF}) \, \left(\frac{16}{\cal C}  \right)^2  
 \left( \frac{10 \, {\rm keV}}{M_{\rm TD}} \right)^5 \left( \frac{m_F}{10^3 \, {\rm GeV}} \right)^4 
\,. 
 \label{lifetime}
\end{equation}

For TD to be a dark matter, its lifetime has to be longer than the age of the Universe, $\sim 10^{17}$ sec, 
which places an upper bound for the TD mass, $M_{\rm TD} \lesssim 10 \, {\rm keV}$~\footnote{
Since $M_{\rm TD} > 2 m_\nu$, 
the $D \to \bar{\nu}\nu$ decay rate might be included in the estimate of $\tau_{\rm TD}$, 
which, however, turns out to be negligible due to the large suppression factor  
$(m_{\nu}/(\alpha_{\rm EM}M_{\rm TD}))^2 \sim 10^{-10}$ compared to the decay rate of Eq.(\ref{decay-width2}). 
}, for the one-family TC model with $m_F=10^3$ GeV. 
This constraint also gives a lower bound on the TD decay constant through Eq.(\ref{PCDC}), 
$F_{\rm TD}\gtrsim10^{11}\gev$, which indeed implies the decoupled TD. 
Recall that, as noted above, the extremely large $F_{\rm TD}$ is possible 
in the {\it extremely walking} case when $\alpha\approx\alpha_{\rm cr}$ which 
is realized by the nonperturbative 
beta function Eq.(\ref{NPbeta}) coupling the TD to the nonperturbative scale anomaly Eq.(\ref{PCDC:original}).

\section{Thermal production}

The TD will be generated at the same time or right after 
the techni-fermion condensate takes place 
at the temperature $T=\mu_{\rm cr}$ where $\alpha = \alpha_{\rm cr}$ and 
$\mu_{\rm cr}$ satisfies $m_F ={\cal O}(10^3\, {\rm GeV}) < \mu_{\rm cr}<  
\Lambda_{\rm TC} (\simeq F_{\rm TD}) = {\cal O}(10^{11}\, {\rm GeV})$. 
As noted above, in addition, due to the Miransky scaling Eq.(\ref{miransky}), 
the dynamical mass $m_F$ should be much 
smaller than the other two scales, so that $m_F/\mu_{\rm cr} \ll 1$ and 
$m_F/\Lambda_{\rm TC} \ll 1$, i.e., $m_F \ll \mu_{\rm cr} < \Lambda_{\rm TC}$. 
This hierarchical structure has actually been confirmed~\cite{Hashimoto:2010nw} 
by an explicit calculation in the walking TC, which predicts 
$m_F/\Lambda_{\rm TC} \simeq 10^{-9}$ and $\mu_{\rm cr}/\Lambda_{\rm TC} \simeq 10^{-3}$ 
for the case with an {\it extremely large scale hierarchy}. 
We thus see that, due to the large decay constant $F_{\rm TD}\gtrsim 10^{11}$ GeV, 
the TD decouples (with the decoupling temperature $T_d \sim 10^{10}$ GeV)  
from the thermal equilibrium as soon as it is produced at $T=\mu_{\rm cr}< F_{\rm TD}$. 
We will take $\Lambda_{\rm TC} (\simeq F_{\rm TD})  =10^{12}\gev$, $\mu_{\rm cr}=10^8$ GeV 
and $m_F\simeq 10^3\gev$ as the reference values inspired by the result of Ref.~\cite{Hashimoto:2010nw}.

Though the TD decouples  from the thermal equilibrium right after its generation, 
it could be produced through scatterings of particles which are 
in the thermal equilibrium at $T \lesssim \mu_{\rm cr}$. 
Those thermal particles are assumed to include techni-hadrons and
techni-fermions with masses of ${\cal O}(m_F)$ as well as the SM particles. 
The most dominant contribution is expected to come from the processes 
including QCD interactions with the relatively large QCD coupling $\alpha_s \sim 0.1$. 
Among the QCD processes, the leading contributions arise from 
those having single TD in the final state with the suppression factor of $1/F_{\rm TD}$. 
Such  processes turn out to be 
$Q(q) + g  \to D + Q(q)$ ($\bar{Q}(\bar{q}) + g  \to D + \bar{Q}(\bar{q})$), 
$Q(q) + \bar{Q}(\bar{q}) \to D + g$, 
$P_c + g \to D + P_c$, $P_c + P_c \to D + g$ and $g + g \to D +g$, 
where $q(Q)$, $g$ and $P_c$ 
denote the SM(techni-) quark, gluon, and colored-techni-pion, 
respectively. 
Since the TD production itself accompanies breaking of the dilatation/scale symmetry, 
those processes should involve the anomalous vertex breaking the dilatation symmetry coupled to 
the QCD gluon field strength $G_{\mu\nu}$, 
${\cal L} \ni - (\beta(\alpha_s)/(2\alpha_s F_{\rm TD})) D \,{\rm tr}[G_{\mu\nu}^2]$. 
At non-trivial leading order of $\alpha_s/F_{\rm TD}$, 
the form of those cross sections then goes like $\sim \alpha_s^3/F_{\rm  TD}^2$.

Plugging those cross sections into the Boltzmann equation, we can estimate 
the relic abundance of TD, $Y_{\rm TD}\equiv n_{\rm TD}/s$, 
the ratio of the number density of TD  to the entropy density. 
The backreaction is safely neglected  
since the TD number density is much smaller than the photon number density in the thermal equilibrium. 
Using the boundary condition $Y_{\rm TD}(T=\mu_{\rm cr})=0$,     
we then find the relic abundance at present $T=T_0$,  
\begin{equation} 
Y_{\rm TD}^{\rm tp}(T_0) \simeq 10^{-5} \left( \frac{\mu_{\rm cr}}{10^8 \,{\rm GeV}}  \right) 
\left( \frac{200}{g_*(\mu_{\rm cr})} \right)^{3/2} \left(\frac{10^{12}\, {\rm GeV}}{F_{\rm TD}} \right)^2
\,, \label{YTD:tp}
\end{equation} 
for the TC one-family model. 
This leads to the thermally produced relic density of TD, 
\begin{eqnarray} 
  \Omega^{\rm tp}_{\rm TD} h^2 
  &\simeq& 10^{-4} \left( \frac{\mu_{\rm cr}}{10^8\,{\rm GeV}}  \right) \left( \frac{M_{\rm TD}}{100\,{\rm eV}} \right) 
\nonumber \\ 
&& 
\times 
\left( \frac{200}{g_*(\mu_{\rm cr})} \right)^{3/2} \left(\frac{10^{12}\, {\rm GeV}}{F_{\rm TD}} \right)^2
\,, 
\label{ther-TD-relic}
\end{eqnarray} 
which is, with {\it a priori} consistency with some astrophysical constraints taken into account, 
much smaller than the observed dark matter density $\simeq 0.1$.

\section{Non-thermal production}

\begin{figure}[t]

\begin{center} 
\includegraphics[scale=0.7]{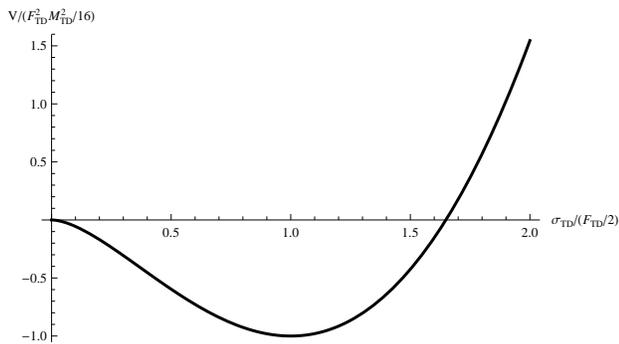}
\end{center}
\caption{The illustration of the TD potential Eq.(\ref{VTD}) 
which has the minimum at $\sigma_D=F_{\rm TD}/2$ with the vacuum energy 
$|V_{\rm min}|=F_{\rm TD}^2M_{\rm TD}^2/16$. } 
\label{potential}
\end{figure}

Analogously to the case of axion dark matter~\cite{Kim:2008hd}, 
the population of TD could be accumulated by ``misalignment"  
of the classical TD field $\sigma_D$ with the scaling dimension $(3-\gamma_m) \simeq 2$. 
Below $T=\mu_{\rm cr}$, the $\sigma_D$  develops a potential 
consistently with the nonperturbative scale anomaly Eq.(\ref{PCDC})~\cite{Miransky:1989qc}: 
\begin{equation} 
 V(\sigma_D) \simeq \frac{F_{\rm TD}^2 M_{\rm TD}^2}{16} \left( \frac{2\sigma_D}{F_{\rm TD}} \right)^2 
 \left[ \log \left(\frac{2 \sigma_D}{F_{\rm TD}}\right)^2 -1  \right]
 \,. \label{VTD}
\end{equation} 
The corresponding form of the potential is depicted in Fig.~\ref{potential}. 
Note that this form is fairly stable against 
both thermal and radiative corrections: 
thermal corrections are not generated since the TD decouples 
from the thermal equilibrium no sooner than its generation, 
while radiative corrections are almost negligible due to the large suppression by $F_{\rm TD}$.

Examining the dynamics at the classical level based on the potential Eq.(\ref{VTD}) as well as 
the scale invariant kinetic term, it turns out~\cite{longer-version} that 
the $\sigma_D$ starts oscillating with the initial position of the order of $F_{\rm TD}$ 
as soon as the TD mass $M_{\rm TD}$ becomes comparable with 
the Hubble parameter $H$ where $H^2 = \frac{\pi^2}{30} g_* T^4/(3M_P^2)$. 
It happens at the temperature $T_{\rm os} \simeq 10^5 \, {\rm GeV} \times (M_{\rm TD}/{\rm keV})^{1/2}
(200/g_*(T_{\rm os}))^{1/4}$, which is much later than the TD generation at $T=\mu_{\rm cr}=10^8$ GeV. 
During the oscillation, 
the number density per comoving volume is preserved due to the adiabatic expansion of the Universe, 
i.e., $ \rho_{\rm TD}(T_0) = \rho_{\rm TD}(T_{\rm os}) \cdot
s(T_0)/s(T_{\rm os})$. 
The non-thermally produced relic density is thus estimated to be  
\begin{eqnarray} 
\Omega_{\rm TD}^{\rm ntp}h^2 
&\simeq&  
11 \left( \frac{\theta_0}{0.1} \right)^2  
\left(\frac{N_{\rm TC}N_{\rm TF}}{16}\right) 
\nonumber \\ 
&&
\times 
 \left(\frac{200}{g_*(T_{\rm os})} \right)  \left(\frac{m_F}{10^3 {\rm GeV}} \right)^4
  \left(\frac{10^5 {\rm GeV}}{T_{\rm os}} \right)^3, \label{non-ther-TD-relic}
\end{eqnarray} 
where $\theta_0=(\sigma_D)_0/(F_{\rm TD}/2)-1$ parametrizes difference between 
the initial position $(\sigma_D)_0$ and $\langle \sigma_D \rangle = F_{\rm TD}/2$  
in the potential and we have used $T_0 = 2.4 \times 10^{-4}$ eV, $g_*(T_0)=43/11$ and 
$\rho_{\rm cr}/h^2 = 0.8 \times 10^{-46}$ GeV$^4$.

\section{Discussion and conclusion}

\begin{figure}
\begin{center}
\includegraphics[scale=0.6]{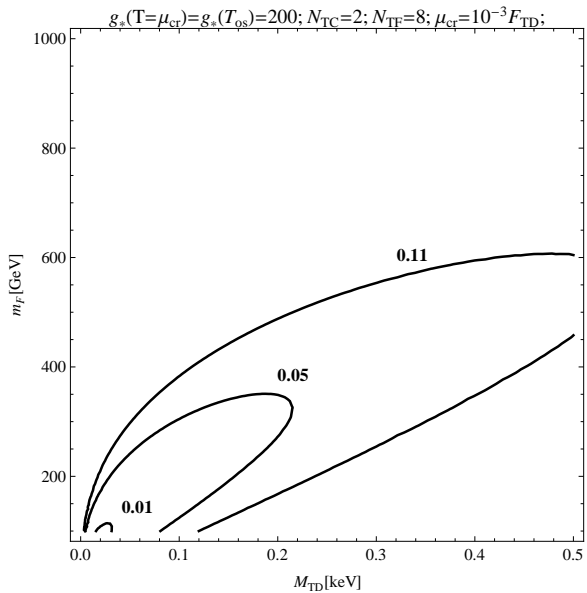}
\end{center}
\caption{The contour plot of the TD relic density 
on $M_{\rm TD}$-$m_F$ plane in the case of one-family model with 
$N_{\rm TC}=2$, $N_{\rm TF}=8$. 
Here $\mu_{\rm cr} = 10^{-3} F_{\rm TD}$ has been taken and the initial position of $\sigma_D$ 
chosen to be $\theta_0=(\sigma_D)_0/(F_{\rm TD}/2) - 1 = 0.1$ as a reference point. } 
\label{NTP-contour} 
\end{figure}

Combining the thermal (\eq{ther-TD-relic}) and  
the non-thermal (\eq{non-ther-TD-relic}) productions, 
we estimate the total relic density of TD as $\Omega_{\rm TD}^{\rm tot} h^2=\Omega_{\rm TD}^{\rm tp} h^2 + \Omega_{\rm TD}^{\rm ntp} h^2$. 
See Fig.~\ref{NTP-contour} which shows a contour plot on $m_F$-$M_{\rm TD}$ plane 
for the one-family model.  
The values of $M_{\rm TD}$ has been restricted to a region, $0.01 \le M_{\rm TD} \le 500$ eV, 
to be consistent with {\it a priori} astrophysical and cosmological constraints discussed below.   
The excluded regions correspond to the domains outside the curve for $\Omega_{\rm TD}^{\rm tot}h^2 = 0.11$, 
which are separated into two areas: one is due to the excessive non-thermal production (upper area), 
while the other one due to the excessive thermal production (lower area).

Some phenomenological bounds on the TD mass or the decay constant   
would be induced from 
cosmological and astrophysical constraints 
on the intrinsic nature of the long-lived decoupled TD. 
Several comments on such an issue are in order.

The TD mass may be constrained by the experimental test of 
the gravitational inverse-square law~\cite{gravity}, 
which gives the lower bound, $M_{\rm TD} \gtrsim 0.01 \ev$, 
with the coupling enhanced by a factor of $(M_P/F_{\rm TD})^2 \sim 10^{16}$ 
compared to a generic dilaton coupling associated with the Planck scale physics. 

The excessive decay of TD with $\kev$ mass into photons would affect 
the observed flux of X-rays in the sky~\cite{Bazzocchi:2008fh}. 
 For the one-family model quoted in Fig.~\ref{NTP-contour}, 
the X-ray background requires the TD mass to be less than around $500 \ev$.

Since TD couples to nucleons, 
the TD decay constant $\FTD$ may be constrained by energy loss in stars 
through the TD production out of stars similarly to the case of axion~\cite{Kim:2008hd}. 
The most severe constraint would then come from neutron star cooling to 
give a lower bound on $F_{\rm TD}$, $\FTD \gtrsim 10^9\gev$. 
One can see, however, that this constraint is well satisfied if the lifetime of TD is long enough 
for TD to be a dark matter 
as in~\eq{lifetime}: indeed, it just places a weaker upper bound on the TD mass, $M_{\rm TD}\lesssim 1\mev$, 
in contrast to the case of axion, 
which is due to the discrepancy between their energy densities, 
namely, $F_{\rm TD}^2M_{\rm TD}^2 \sim m_F^4$ and $f_a^2 m_a^2 \sim \Lambda_{\rm QCD}^4$.

In conclusion, 
the population of TDs with mass range between around 0.01 eV and  500 eV  
can be produced in the early Universe, 
dominantly through the non-thermal production, 
which is rich enough to be the main component of dark matter (See Fig.~\ref{NTP-contour}).

The relic TD may be detected through the monoenergetic X-rays from the sky,  
where the source for such detection will be provided by the two-photon decay of TD or 
the electromagnetic resonant cavity based on the TD-photon conversion  
in strong electromagnetic background~\cite{Cho:2007cy}. 
This issue is to be explored in detail in  the future publication~\cite{longer-version}.

\section*{Acknowledgments}

 We would like to thank H.~S.~Fukano, 
J.~E. Kim, B. Kyae, and K. Yamawaki for useful comments. 
This work is supported by the Korea Research 
Foundation Grant funded by the Korean Government
(KRF-2008-341-C00008).  


\begin{thebibliography}{99}



\bibitem{Weinberg:1975gm}
  S.~Weinberg,
  Phys.\ Rev.\ D {\bf 13}, 974 (1976);
  L.~Susskind,
  Phys.\ Rev.\ D {\bf 20}, 2619 (1979).


\bibitem{Farhi:1980xs}
 For reviews, see, e.g.,  E.~Farhi and L.~Susskind,
  Phys.\ Rept.\  {\bf 74}, 277 (1981);
  K.~Yamawaki,
  Lecture at 14th  Symposium on Theoretical Physics, Cheju, Korea,  July 1995, 
  arXiv:hep-ph/9603293;
  C.~T.~Hill and E.~H.~Simmons,
  Phys.\ Rept.\  {\bf 381}, 235 (2003)
  [Erratum-ibid.\  {\bf 390}, 553 (2004)]; 
  F.~Sannino,
  Acta Phys.\ Polon.\  {\bf B40}, 3533-3743 (2009). 

\bibitem{Holdom:1981rm}
  B.~Holdom,
  Phys.\ Rev.\  D {\bf 24} (1981) 1441.

\bibitem{Yamawaki:1985zg}
  K.~Yamawaki, M.~Bando and K.~Matumoto,
  Phys.\ Rev.\ Lett.\  {\bf 56}, 1335 (1986);
  M.~Bando, K.~Matumoto and K.~Yamawaki,
  Phys.\ Lett.\  B {\bf 178}, 308 (1986);
  M.~Bando, T.~Morozumi, H.~So and K.~Yamawaki,
  Phys.\ Rev.\ Lett.\ 
  {\bf 59}, 389 (1987).
  

 

\bibitem{Akiba:1985rr}
  T.~Akiba and T.~Yanagida,
  Phys.\ Lett.\ B {\bf 169}, 432 (1986); 
  T.~W.~Appelquist, D.~Karabali and L.~C.~R.~Wijewardhana,
  Phys.\ Rev.\ Lett.\  {\bf 57}, 957 (1986);
  T.~Appelquist and L.~C.~R.~Wijewardhana,
  Phys.\ Rev.\ D {\bf 36}, 568 (1987).
  




\bibitem{Hong:2004td}
  D.~K.~Hong, S.~D.~H.~Hsu and F.~Sannino,
  Phys.\ Lett.\  B {\bf 597}, 89 (2004);
  F.~Sannino and K.~Tuominen,
  Phys.\ Rev.\  D {\bf 71}, 051901 (2005).
  
  
\bibitem{Dimopoulos:1979es}
  S.~Dimopoulos and L.~Susskind,
  Nucl.\ Phys.\  B {\bf 155}, 237 (1979); 
  E.~Eichten and K.~D.~Lane,
  Phys.\ Lett.\  B {\bf 90}, 125 (1980).
  

\bibitem{Miransky:1984ef}
  V.~A.~Miransky,
  Nuovo Cim.\  A {\bf 90}, 149 (1985).

\bibitem{Hong:1989zza}
  D.~K.~Hong and S.~G.~Rajeev,
  Phys.\ Lett.\  B {\bf 240}, 471 (1990).

\bibitem{Kaplan:2009kr}
  D.~B.~Kaplan, J.~W.~Lee, D.~T.~Son and M.~A.~Stephanov,
  Phys.\ Rev.\  D {\bf 80}, 125005 (2009). 
  
  
  
  
\bibitem{Miransky:1996pd}
  V.~A.~Miransky and K.~Yamawaki,
  Phys.\ Rev.\ D {\bf 55}, 5051 (1997)
  [Erratum-ibid.\ D {\bf 56}, 3768 (1997)].
  
  



\bibitem{Bardeen:1985sm}
  W.~A.~Bardeen, C.~N.~Leung and S.~T.~Love,
  Phys. Rev. Lett. {\bf 56}, 1230 (1986);
  Nucl.\ Phys.\  B {\bf 273} (1986) 649.









\bibitem{Caswell:1974gg}
 W.~E.~Caswell,
 Phys.\ Rev.\ Lett.\  {\bf 33}, 244 (1974);
 T.~Banks and A.~Zaks,
 Nucl.\ Phys.\  B {\bf 196}, 189 (1982).



\bibitem{Hashimoto:2010nw}
  M.~Hashimoto and K.~Yamawaki,
  Phys.\ Rev.\  D {\bf 83}, 015008 (2011).










\bibitem{Shuto:1989te}
  S.~Shuto, M.~Tanabashi and K.~Yamawaki,
 in {\it Proc. 1989 Workshop on Dynamical Symmetry Breaking}, 
      Dec. 21-23, 1989, Nagoya, eds. T. Muta and K. Yamawaki 
      (Nagoya Univ., Nagoya, 1990) 115-123.   



\bibitem{Bardeen:1991sv}
  W.~A.~Bardeen, S.~T.~Love,
  Phys.\ Rev.\  {\bf D45}, 4672-4680 (1992). 




\bibitem{Carena:1992cg}
  M.~S.~Carena and C.~E.~M.~Wagner,
  Phys.\ Lett.\  B {\bf 285}, 277 (1992);
  M.~Hashimoto,
  Phys.\ Lett.\  B {\bf 441}, 389 (1998).  
  
  
  
  





  

\bibitem{Harada:2003dc}
  M.~Harada, M.~Kurachi and K.~Yamawaki,
  Phys.\ Rev.\ D {\bf 68}, 076001 (2003);
  M.~Kurachi and R.~Shrock,
  JHEP {\bf 0612}, 034 (2006). 





\bibitem{Haba:2010hu}
  K.~Haba, S.~Matsuzaki and K.~Yamawaki,
  Phys.\ Rev.\  D {\bf 82} (2010) 055007.
  
  
  

\bibitem{Manohar:1983md}
  A.~Manohar and H.~Georgi,
  Nucl.\ Phys.\  B {\bf 234}, 189 (1984);
  H.~Georgi and L.~Randall,
  Nucl.\ Phys.\  B {\bf 276}, 241 (1986);
  H.~Georgi,
  Phys.\ Lett.\  B {\bf 298}, 187 (1993). 





\bibitem{Appelquist:2010gy}
  T.~Appelquist and Y.~Bai,
  Phys.\ Rev.\  D {\bf 82}, 071701 (2010). 



  
\bibitem{Miransky:1989qc}
  V.~A.~Miransky and V.~P.~Gusynin,
  Prog.\ Theor.\ Phys.\  {\bf 81}, 426 (1989).
  
  
  
  




\bibitem{Nonoyama:1989dq}
  T.~Nonoyama, T.~B.~Suzuki, K.~Yamawaki,
  Prog.\ Theor.\ Phys.\  {\bf 81}, 1238 (1989).   
  

\bibitem{Leung:1989hw}
  C.~N.~Leung, S.~T.~Love, W.~A.~Bardeen,
  Nucl.\ Phys.\  {\bf B323}, 493 (1989);
  K.~-i.~Kondo, H.~Mino, K.~Yamawaki,
  Phys.\ Rev.\  {\bf D39}, 2430 (1989); 
  T.~Nonoyama, T.~B.~Suzuki, K.~Yamawaki,
  Prog.\ Theor.\ Phys.\  {\bf 81}, 1238 (1989).
  






\bibitem{Hashimoto:2011ma}
  M.~Hashimoto,
  Phys.\ Rev.\  {\bf D83}, 096003 (2011); 
  S.~Matsuzaki, K.~Yamawaki,
  [arXiv:1109.5448 [hep-ph]].




\bibitem{longer-version} 
K.~-Y.~Choi, D.~K.~Hong and S.~Matsuzaki, in preparation. 














  
\bibitem{Kim:2008hd}
  J.~E.~Kim and G.~Carosi,
  Rev.\ Mod.\ Phys.\  {\bf 82} (2010) 557. 














\bibitem{gravity}
  E.~Fischbach, D.~E.~Krause, V.~M.~Mostepanenko, M.~Novello,
  Phys.\ Rev.\  {\bf D64 } (2001)  075010.
  S.~Dimopoulos, A.~A.~Geraci,
  Phys.\ Rev.\  {\bf D68 } (2003)  124021.
  A.~A.~Geraci, S.~J.~Smullin, D.~M.~Weld, J.~Chiaverini, A.~Kapitulnik,
  Phys.\ Rev.\  {\bf D78 } (2008)  022002.
  A.~A.~Geraci, S.~B.~Papp, J.~Kitching,
  [arXiv:1006.0261 [hep-ph]].





\bibitem{Bazzocchi:2008fh}
  F.~Bazzocchi, M.~Lattanzi, S.~Riemer-Sorensen, J.~W.~F.~Valle,
  JCAP {\bf 0808 } (2008)  013.



\bibitem{Cho:2007cy}
  Y.~M.~Cho and J.~H.~Kim,
  Phys.\ Rev.\  D {\bf 79} (2009) 023504. 









\end{thebibliography}
\end{document}